\documentstyle[11pt,fleqn]{article}

\begin{document}

\baselineskip=18pt
\begin{center}
{\Large \bf Fermi Gas in  Harmonic oscillator potentials
}\\
\ \ \\
 X.X. Yi$^{1,3}$ \ \ \ J.C.Su$^2$ \\
\end{center}
\begin{center}
$^1${\it Institute of Theoretical Physics, Northeast Normal University,\\
Changchun 130024, China\\}
$^2${\it Department of Physics, Jilin University,\\
Changchun 130023, China\\}
$^3${\it Institute of Theoretical Physics, Academia Sinica, Peking 100080, China\\}
{\large Abstract}
\end{center}

Assuming the validity of grand canonical statistics,
we study the properties of a spin-polarized Fermi gas in
harmonic traps. Universal forms of Fermi temperature $T_F$, internal energy
$U$ and the specific heat per particle of the trapped Fermi gas are calculated
as a {\it function} of particle number, and the results compared with those of
infinite number particles.
\vspace{2cm}

The ideal Fermi gas is an old and
well-understood problem since the non-interaction Fermi gas  is a good
zeroth-order approximation for many familiar systems.
Like the trapped degenerate atomic gases [1-3] that
provide exciting opportunities for the manipulation and quantitative study of
quantum statistical effects, the behavior of trapped Fermi gas also
merits attention, both as a degenerate quantum system in its own right
and as a possible precursor to a paired Fermi condensate at low
temperature[4] though it perhaps is not as dramatic as the phase transition
associated with bosons.

The trapped atomic gases reported in Ref.[1-3] are dilute. The effects of
predominantly short-range atom-atom interactions are
therefore week. For dilute spin-polarized Fermi gases, the s-wave scattering
amplitude vanishes due to the antisymmetry of the many-Fermi wave function,
and the p-wave scattering is small at low energy. Therefore the dilute Fermi
gas is usually treated as an ideal Fermi gas.

Unlike the usual system in most textbooks of statistical mechanics, e.g.[5],
however, the number of trapped atoms in Ref.[1-3] is finite (about $10^9$).
For finite number of trapped fermions,
{\it the applicability of the usual thermodynamical calculations
and the neglect of the ground energy are no longer available}.
For these relatively
 low numbers (compared to $10^{23}$), the effects caused by the above
 approximation are nonvanishing.

 Harmonic traps provide a particularly simple realization of the
 confined Fermi system.
In this trap, Butts and Rokhsar[6] calculated the spatial and momentum
distributions for trapped fermions using Thomas-Fermi approximation.
The spatial distribution of the trapped cloud provides an explicit
visualization of a real-space "Fermi sea" and the momentum distribution,
unlike the spatial one, is isotropic.

In this paper, we calculate the Fermi temperature, internal energy and
the specific heat per particle of the finite number trapped Fermi gas.
We indeed find marked differences from the usual treatments:
a correction to the Fermi temperature, internal energy and the specific
heat per particle of the finite number trapped fermions,
a constraint to the number of trapped fermions for given trap
frequencies.

 Consider $N$ spin-polarized fermions of mass $m$ moving in anisotropic
 oscillator potentials. The single-particle levels are familiar
\begin{equation}
E_N=E(n_x,n_y,n_z)=\sum_{i=x,y,z}\hbar\omega_i(n_i+\frac 1 2),
\end{equation}
where $\omega_i(i=x,y,z) $is the trap frequencies in the $i$ direction,
$n_i$ is non-negative integer.
{\it Note that each oscillator state is assumed
to be filled with a single fermion, since only one spin orientation
is confined by the magnetic trap.}
Under real
experimental conditions of trapped atomic gases,
the temperature is high on the scale of
the trap level spacing, namely, $k_BT>>\hbar\omega_i(i=x,y,z).$ Therefore,
within the canonical ensemble, the partition function
\begin{equation}
Q(\beta)=\sum_N e^{-\beta E_N}=\sum_{n_x,n_y,n_z=0}^{\infty}
e^{-\beta(n_x\omega_x+n_y\omega_y
+n_z\omega_z)}=\prod_{i=x,y,z}\frac{1}{1-e^{-\beta\omega_i}}
\end{equation}
of such a system without interactions can be expanded as follows:
\begin{equation}
Q(\beta)\simeq a_2\beta^{-3}+a_1\beta^{-2}+a_0\beta^{-1}+a_{-1}+O(\beta)
\end{equation}
with
\begin{eqnarray}
a_2&=&\frac{1}{\omega_x\omega_y\omega_z}\nonumber\\
a_1&=&\frac 1 2 a_2(\omega_x+\omega_y+\omega_z)\\
a_0&=&\frac {1}{12}a_2(\omega_x^2+\omega_y^2+\omega_z^2+
3\omega_x\omega_y+3\omega_y\omega_z+3\omega_z\omega_z)\nonumber\\
a_{-1}&=&\frac 1 8 +\frac {1}{24}(\omega_x/\omega_y+\omega_x/\omega_z+
\omega_y/\omega_z+\omega_y/\omega_x+\omega_z/\omega_x+\omega_z/\omega_y),
\end{eqnarray}
where, $\beta=k_BT$, $k_B$ is the Boltzmann constant, and the contribution
of the ground state was singled away for special treatment.
 On the other hand, by making  use of
 \begin{equation}
Q(\beta)=\sum_Ne^{-\beta E_N}=\int_{\varepsilon_0}
^{\infty}e^{-\beta E}\rho(E)dE
\end{equation}
the partition function can be calculated equivalently,
where $\rho(E)$ is the density of states, $\varepsilon_0$ stands
for the energy of the ground state. Comparing with eq.(3), it
is proved that the density of states $\rho(E)$ takes a form:
\begin{equation}
\rho(E)=b_0+b_1E+b_2E^2+...   .
\end{equation}
This expansion can't contain the terms $E^r$ with $r$ being negative
or non-integer, since they become zero after comparison with
the direct calculation given by eq.(3).
Substitution eq.(7) into eq.(6), one can easily find:
\begin{equation}
Q(\beta)=(b_0+b_1\varepsilon_0+b_2\varepsilon_0^2)\beta^{-1}+
(b_1-2b_2\varepsilon_0)\beta^{-2}+2b_2\beta^{-3}+...
\end{equation}
comparison of eq.(8) with the direct calculation(3) shows that
\begin{equation}
b_0=a_0-\varepsilon_0a_1-\varepsilon^2_0a_2, b_1=a_1+\varepsilon_0a_2,
b_2=0.5a_2
\end{equation}
There are two additional  terms  $b_0$ and $b_1E$
in the density of states in
comparison with the result of Butts[6], as we see in following, these
additional terms will result in the shift of the
thermodynamic quantities.

Now, let us consider a system of $N$ noninteracting fermions
 that the population
$N(E_i)$ of a state with energy $E_i$ is given by the Fermi-Dirac
distribution
\begin{equation}
N(E_i)=\frac{1}{e^{\beta(E_i-\mu)}+1}
\end{equation}
Here, we set the statistical weights corresponding the state $E_i$, $g_i=1$.
$\mu$ stands for the chemical potential, which is determined by the
constraint that the total number of the particles in the system is $N$:
\begin{equation}
           N=\sum_i^{\infty}N(E_i).
\end{equation}
At zero temperature the Fermi-Dirac distribution factor is unity for
energies less than the Fermi energy $E_F=\mu(T=0,N)$,
and zero otherwise. A straightforward integration of Eq.(11) gives:
\begin{equation}
E_F^3+\frac{3b_1}{2b_2}E_F^2+3\frac{b_0}{b_2}E_F-\frac{3}{b_2}N^{'}=0
\end{equation}
with $N^{'}=N+\frac 1 3 b_2\varepsilon_0^2+\frac 1 2 b_1\varepsilon_0^2+
b_0\varepsilon_0$.
The terms with $b_0$ and $b_1$ are corrections from finite number effects to
Fermi energy. In the large number particle limit, $E_F=(\frac{3}{b_2}N^{'})
^{1/3}$. This  just is the result of Ref.[6,7].
{\it Some words of caution are now in order.
Fermi energy $E_F$ is determined by eq.(12),
i.e. $E_F$ is a solution of eq.(12).
From the mathematical point of view, the eq.(12)
has a only real solution
\begin{equation}
E_F-(\frac{3}{b_2}N^{'})^{1/3}=-\frac 1 4 \frac{b_1}{b_2}-
\frac 1 2 \frac {b_0}{b_2}
\end{equation}
in the situation of the number of trapped fermior satisfying
\begin{equation}
N^{'}<N_{max}, N_{max}=[-(\frac p 3)^{3/2}+q_0]b_0/3,
\end{equation}
where
\begin{equation}
p=-\frac{3b_1^2}{4b_2^2}+3\frac{b_0}{b_2},q_0=\frac{b_1^3}{4b_2^3}-
\frac{3b_0b_1}{2b_2^2}.
\end{equation}
Otherwise, the eq.(12) have three real solution, but only the
positive smallest one meets the requirements for Fermi energy
from the physical point of view. We would like to point out that
the results presented here are
valid for $k_BT>>\hbar\omega_i(i=x,y,z).$ If this condition is broken,
we should make use of a numerical method[8] to study the effects of finite
particles.}

For finite temperature, neglecting the effect of zero-point energy, the eq.(11)
gives:
\begin{equation}
N=b_2k_B^3T^3f_3(z)+b_1k_B^2T^2f_2(z)+b_0k_BTf_1(z)
\end{equation}
where $f_n(z)$ stands for the Fermi integral, which is given that
$$
f_n(z)=\frac{1}{\Gamma(n)}\int_0^{\infty}\frac{x^{n-1}}{z^{-1}x^n+1}dx,
\Gamma(n)=\int_0^{\infty}e^{-x}x^{n-1}dx.
$$
$z$ is a fugacity.
The last two terms in eq.(16) are due to the effects of finite particles.
If one introduces a temperature
\begin{equation}
T_F^0=\frac{1}{k_B}(\frac {N}{b_2f_3(z_F)})^{1/3},
\end{equation}
which denotes the Fermi  temperature of infinite number fermions
trapped in anisotropic oscillator potentials[6], i.e. $k_BT_F^0=\mu$.
Then the Fermi temperature $T_F$
{\it defined as the same as in most textbooks}
for finite number  fermions in the case
of $\varepsilon_0=0$ takes
a form
\begin{equation}
T_F=T_F^0[1-\frac 1 3 \frac{b_1}{b_2}\frac{1}{k_BT_F^0}
\frac{f_2(z_F)}{f_3(z_F)}-
\frac 2 3 \frac{b_0}{b_2}\frac{1}{(k_B^2T_F^0)^2}\frac{f_1(z_F)}{f_3(z_F)}],
\end{equation}
where $z_F=z|_{T=T_F^0}$.
The last two terms in eq.(18)
are correction to Fermi energy from the finite number effect,
it shows that the Fermi temperature for finite trapped fermions
is lower than that in the case [7] of infinite trapped fermions.

Generally speaking, the specific heat is more interest from the experimental
point of view, since the low-temperature behavior of specific heat $c$
is generally treated as the hallmark of onset of phase transition .
It is well known that
the specific heat can be derived from the internal energy, which is
expressed by
\begin{equation}
\beta U=\beta\int_0^{\infty}
\frac{E\rho(E)}{exp[\beta(E-\varepsilon_0)]+1}dE
\end{equation}
Here we dropped the  ground state energy, which is less than
$2\varepsilon_0$ and remains constant. Substituting eq.(7)
into eq.(19), one easily finds
\begin{equation}
U=\frac{b_2}{\beta^4}f_4(z)+\frac{b_1}{\beta^3}f_3(z)+\frac{b_0}{\beta^2}
f_2(z),
\end{equation}
the first term remains unchanged in the limit of
very large number particles, while the last two terms
close to zero.
The specific heat per particle of the trapped fermions is given that[6]
\begin{equation}
c=[\frac 1 N \frac{\partial U}{\partial T}]_N,
\end{equation}
straightforward calculation gives
\begin{eqnarray}
c&=&\frac 1 N\{ 4b_2k_B^4T^3f_4(z)+3b_1k_B^3T^2f_3(z)+2b_0k_B^2Tf_2(z)-\nonumber\\
& \ &\frac 3 2 b_2k_B^4T^3\frac{f_{3/2}(z)f_3(z)}{f_{1/2}(z)}-
\frac 3 2 b_1k_B^3T^2\frac{f_{3/2}(z)f_2(z)}{f_{1/2}(z)}-
\frac 3 2 b_0k_B^2T\frac{f_{3/2}(z)f_1(z)}{f_{1/2}(z)}
\}
\end{eqnarray}
It is evident that the specific heat per particle has increased
due to the finite particle effects. The results are  illustrated in Fig.1.
As state in most textbooks of statistical mechanics, for ideal Fermi gas in a
3D box, the specific heat increases proportionally with temperature $T$
for $T<<T_F$, whereas it closes to a constant $1.5k_B$ for $T\rightarrow \infty$.
For trapped fermions, however, the specific heat increases predominantly
with $T$ for $T<<T_F$, and close to $3k_B$ with $T\rightarrow \infty$.
{\it Physically, the difference between the case of 3D box and that of trapped
fermions is the confinment.This leads to the difference in their energy levels.
Consequently, different energy levels would result in different dependence
of specific heat on temperature.}

In conclusion, we have discussed the behaviors of fermions
trapped in an anisotropic oscillator potentials, it was shown that
corrections due to the effect of finite fermions and
the ground state energy are small, whether they are observable depends on
the experimental parameters. We believe the results will be
useful to them who is working experimentally in this area.\\
{\bf ACKNOWLEDGMENTS:}\\
One of us(X.X.Yi) would like to thank prof. C.P.Sun for
his helpful discussions.

\newpage
{\it Fig.1 The specific heat per particle vs. temperature. The dots
show the situation for a very large number of particles($N=10^{23}$),
and the dashed line that for a smaller number($10^{8}$).
The trap frequencies are $\omega_x=500Hz,\omega_y=600Hz,\omega_z=800Hz.$}\\

\end{document}